\documentstyle[epsfig,12pt]{article}

\begin{document}

\vspace*{-15mm}

\begin{center}
  {\Large {\bf Topologically Massive Gauge Theory:
               Wu-Yang Type Solutions }} \\[5mm] 
  {\large K. Saygili\footnote{Electronic address: ksaygili@yeditepe.edu.tr}}
  \\[3mm]
  {Department of Mathematics, Yeditepe University,} \\ 
  {Kayisdagi, 34755 Istanbul, Turkey} \\[10mm]

  Abstract: \\

\end{center}

   We discuss Wu-Yang type solutions of Maxwell-Chern-Simons and 
Yang-Mills-Chern-Simons theories. There exists a natural scale of length 
which is determined by the inverse topological mass $\nu \sim ng^{2}$. 
We obtain the non-abelian solution by means of a $SU(2)$ gauge transformation 
of Dirac magnetic monopole type solution. In the abelian case, field strength 
locally determines the gauge potential up to a closed term via self-duality 
equation. We introduce a transformation of the gauge potential using dual 
field strength which can be identified with the gauge transformation in the 
abelian solution. Then we present Hopf map $S^{3} \longrightarrow S^{2}$ 
including the topological mass. This leads to a reduction of the field 
equation onto $S^{2}$ using local sections of $S^{3}$. The local solutions 
possess a composite structure consisting of both magnetic and electric 
charges. These naturally lead to topologically massive Wu-Yang solution 
which is based on patching up the local potentials by means of a gauge 
transformation. We also discuss solutions with different first Chern 
numbers. There exist a fundamental scale $\lambda=2\pi/g^{2}$ over which 
the gauge function is single-valued and periodic for any integer $n$ in 
addition to the fact that it has a smaller period $\lambda/n$. We also 
discuss Dirac quantization condition. We present a stereographic view 
of the fibres in the Hopf map. Meanwhile Archimedes map yields a simple 
geometric picture for the Wu-Yang solution. We also discuss holonomy of 
the gauge potential and the dual-field on $S^{2}$. Finally we point out 
a naive identification of the natural length scale introduced by the 
topological mass with Hall resistivity.

\newpage

\section{Introduction}

  Topologically massive gravity and gauge theory are dynamical theories 
which are specific to three dimensions \cite{DJTS1, DJTS2}, \cite{DJTS3}. 
They are qualitatively different from the Einstein gravity and the 
Yang-Mills gauge theory beside their mathematical elegance and 
consistency. 

  The most distinctive feature of the topologically massive gauge theories 
is the existence of a natural scale of length introduced by the topological 
mass: $[\nu]=[g]^{2}=L^{-1}$ \cite{D}, \cite{J}, in geometric units 
($h=1$, $c=1$). The Dirac \textit{monopole} \cite{GoddardOlive} type 
solution of the Maxwell-Chern-Simons electrodynamics in a space with 
euclidean signature is an example of the essential new feature introduced 
by the topological mass \cite{ANS}. This is a physical system which is a 
priori unrelated to gravity but which nevertheless requires curved space 
for its existence \cite{ANS}.

   We discuss euclidean topologically massive Wu-Yang type solutions 
\cite{WuYang1, WuYang2}, \cite{Ryder} of the Maxwell-Chern-Simons (MCS) 
and the Yang-Mills-Chern-Simons (YMCS) theories on de Sitter (dS) space 
$S^{3}$. 

   In section 2 we present, after a brief account of the YMCS theory, 
the geometric frame including the natural scale of length which is 
determined by the topological mass. We use an intrinsic arclength 
parameterization. This endows us with a dimensionally consistent 
framework for our solutions in geometrical terms. We obtain 
the non-abelian solution by embedding the Dirac solution 
\cite{ANS} into the YMCS theory by means of a $SU(2)$ 
gauge transformation \cite{WuYang1, WuYang2}, \cite{Ryder}. 

   We return to the MCS theory in section 3. In the topologically 
massive electrodynamics, field strength locally determines the gauge 
potential up to a closed term via self-duality equation \cite{TPN}, 
\cite{Deser1}, \cite{WuYang3}. We introduce a transformation of the 
gauge potential using the dual-field strength. This will lead to the 
gauge function in Wu-Yang solution in terms of electric charge.

  Then we present Hopf map $S^{3} \longrightarrow S^{2}$ including 
the topological mass \cite{M, R, T}. This leads to a reduction of 
the field equation onto $S^{2}$ using local sections of $S^{3}$. 
The Hopf map yields local gauge potentials on $S^{2}$ which differ 
by a  gauge transformation. This gives rise to topologically massive 
Wu-Yang solution. The gauge functions, in both the abelian and 
non-abelian cases, contain the topological mass $\nu \sim ng^{2}$ 
which yields the strength $\nu/g \sim ng$ for potentials. This is 
analogous to the non-massive Wu-Yang monopole which consist of the 
dimensionless factor $eg$.

  The local solutions have a composite structure consisting of both 
magnetic and electric charges which are related via integral of the 
field equation. The gauge function can equivalently be given in terms 
of either the magnetic charge or the electric charge. Conversely the 
integrated field equation naturally gives rise to Wu-Yang type solutions 
on $S^{2}$ \cite{WuYang1, WuYang2}, \cite{Ryder} and to loop-variables 
\cite{Mandelstam, Wilson}. We also consider solutions with different 
First Chern numbers.

  In section 4, we present three geometric structures which arise in 
connection with our solutions. We briefly present a stereographic 
view of the fibres in the Hopf map \cite{Pen}, \cite{Berger, Thurston}. 
Meanwhile Archimedes map \cite{MdS}, \cite{AF} provides a simple 
geometric picture for the Wu-Yang solution on $S^{2}$. We also express 
geometric phase \cite{Berry}, \cite{Simon}, \cite{Zumino, Kritsis} 
suffered by a vector upon parallel transport on $S^{2}$ \cite{AF}, 
\cite{Nakahara} in terms of holonomy of the gauge potential or the 
dual-field  \cite{Hannay, AnanStod, Li, BFH}, \cite{A}, \cite{Mostafazadeh}. 

  Then we discuss quantization of the topological mass: $\nu= ng^{2}$ 
and the length scale determined by it in section 5. There exist a 
fundamental scale $\lambda=2\pi/g^{2}$ over which the gauge function 
is single-valued and periodic for any integer $n$ in addition to the 
fact that it has a smaller period $\lambda/n$. We discuss the First 
Chern number and the quantization of the topological mass in terms of 
winding numbers of maps of circles. We also discuss Dirac quantization 
condition.

  The geometric framework presented here might provide a naive explanation 
for the integer quantum Hall effect if the Hall resistance could be 
identified with the natural scale of length introduced by the topological mass.

  The literature consists of various discussions on quantization of the 
topological mass and topological aspects of Wu-Yang type constructions 
(with no solution) \cite{Alvarez, Polychronakos, HenTei, HorvathyNash}, 
beside different monopole/instanton type solutions \cite{Pisarski, AHPS}, 
\cite{PiJackiw, TSH}. However there arise conflicting arguments, see for 
example \cite{Polychronakos, HenTei}, because the essential new feature 
introduced by the topological mass is ignored in topological discussions,
except a few remarks \cite{PiJackiw, TSH}. To the knowledge of the author, 
the solutions and geometric constructions presented here which clarify this 
issue are lacking in the literature .

  A lorentzian version of these solutions and structures on AdS space was 
given by the same author \cite{Saygili1}. The connection of the MCS theory 
with contact geometry and curl transformation was discussed in \cite{Saygili2}.

\section{The Non-Abelian Gauge Theory}

\subsection{Yang-Mills-Chern-Simons Theory}

  The topologically massive $SU(2)$ YMCS theory is given by the 
dimensionless action

\begin{eqnarray} \label{YMCSaction}
S_{YMCS} &=& S_{YM}+S_{CS} \\
&=& -\frac{1}{2} \left[ \int tr \bigg( F \wedge *F \bigg) 
- \nu \int tr \left( F \wedge A 
+ \frac{1}{3} g A \wedge A \wedge A \right) \right] , \nonumber 
\end{eqnarray}

\noindent where $\nu$ is the topological mass. We include the factor 
containing the gauge coupling constant $g$ in (\ref{YMCSaction}) in 
expressions for the field and the gauge potential because of underlying 
geometric reasons that will become clear in advance. This will lead to 
strength $g$ for the potential upon assuming quantization of the 
topological mass. The YMCS action (\ref{YMCSaction}) yields the 
field equation 
 
\begin{eqnarray} \label{YMCSfieldequation}
D*F - \nu F = 0 ,
\end{eqnarray}

\noindent  where $D$ is the gauge covariant exterior derivative. 
The field $2$-form also satisfies the Bianchi identity
 
\begin{eqnarray} \label{Bianchi}
DF=0 .
\end{eqnarray}

\noindent The covariant exterior derivative $D$ respectively
acts as $D*F= \left( d-g\{A, \,\, \} \right)*F$ and 
$DF= \left( d-g[A, \,\, ] \right) F$ on the dual field $1$-form 
$*F$ and the field $2$-form $F$. Here $\{ \,\, , \,\, \}$ denotes 
the anti-commutator and $[ \,\, , \,\, ]$ the commutator. 

  Action (\ref{YMCSaction}) is not invariant under non-abelian 
large gauge transformations

\begin{eqnarray} \label{gaugetransformation}
A'=U^{-1}AU-\frac{1}{g}U^{-1}dU .
\end{eqnarray}

\noindent It changes by 

\begin{eqnarray} \label{Windingnumber}
W=-8\pi^{2}\frac{\nu}{g^{2}} \, w ,
\end{eqnarray}

\noindent neglecting a surface term that vanishes under suitable 
asymptotic convergence conditions on $U$ \cite{DJTS1, DJTS2}, 
\cite{DJTS3}. Here $w$ is the integer winding number

\begin{eqnarray} \label{windingnumber}
w=\frac{1}{48\pi^{2}} 
\int tr \left( U^{-1}dU \wedge U^{-1}dU \wedge U^{-1}dU \right) ,
\end{eqnarray}

\noindent of the gauge transformation \cite{DJTS1, DJTS2}, 
\cite{DJTS3}. Therefore the large gauge transformations, 
which are labeled by the winding number $w$, have a non-trivial 
contribution to the action. 

    In the lorentzian case if we demand the expression $\exp(iS)$ 
to be gauge invariant in order to have a well-defined quantum theory 
via path-integrals, then the change (\ref{Windingnumber}) can be tolerated 
if the topological mass is quantized as

\begin{eqnarray} \label{quantizationoftopologicalmass}
\nu=\frac{1}{4\pi} ng^{2} ,
\end{eqnarray}

\noindent \cite{DJTS1, DJTS2}, \cite{DJTS3}.

   We remark that the action, the field equation and the solutions 
presented here, which are defined on a curved space with euclidean 
signature, are all real valued. We shall refer the relation: 
$\nu \sim ng^{2}$ as the quantization of the topological mass no 
matter if $n$ is an integer until section 5. Because this relation 
will naturally arise in our solutions. 

\subsection{The Natural Scale of Length}

  We shall consider the MCS and YMCS theory over a space with the 
co-frame consisting of the \textit{modified} left-invariant basis 
$1$-forms of Bianchi type $IX$ in Euler parameters \cite{ANS}. The 
topological mass introduces a geometric scale of length. We shall 
use an intrinsic arclength parameterization where the arclength 
parameters are independent of the length scale introduced by the 
topological mass. 

  We scale the unmodified co-frame with the dimensionful factor 
$1/\nu$. This yields 

\begin{eqnarray} \label{coframe}
\omega^{1} &=& -\sin(\nu\psi)d\theta+\cos(\nu\psi)\sin(\nu\theta)d\phi
\nonumber \\
\omega^{2} &=& \cos(\nu\psi)d\theta+\sin(\nu\psi)\sin(\nu\theta)d\phi \\
\omega^{3} &=& d\psi+\cos(\nu\theta)d\phi , \nonumber 
\end{eqnarray}

\noindent in terms of the intrinsic (half) arclength coordinates 

\begin{eqnarray} \label{changeofvariablesS3}
& & \theta=\frac{1}{2}R\tilde{\theta} \hspace*{5mm} , \hspace*{5mm}
\phi=\frac{1}{2}R\tilde{\phi} \hspace*{5mm} , \hspace*{5mm}
\psi=\frac{1}{2}R\tilde{\psi}  \\
& & \hspace*{3mm} =\frac{1}{\nu} \tilde{\theta} \hspace*{17mm} 
=\frac{1}{\nu} \tilde{\phi} \hspace*{17mm}
=\frac{1}{\nu} \tilde{\psi} , \nonumber 
\end{eqnarray}

\noindent which have the dimension of length: 
$[\theta]=[\phi]=[\psi]=L$. This amounts to scaling the 
Cartan-Killing metric by the factor $1/\nu^{2}$

\begin{eqnarray} \label{metricS3}
& & ds^{2} = \eta_{ab}\omega^{a}\omega^{b} \hspace*{10mm} 
\eta_{ab}=diag(1, 1, 1) \\
& & \hspace*{7mm} = d\theta^{2}+d\phi^{2}+2\cos(\nu\theta)d\phi d\psi
+d\psi^{2} \nonumber \\
& & \hspace*{7mm} =d\theta^{2}+\sin^{2} \left ( \nu\theta \right ) d\phi^{2}
+[d\psi+\cos(\nu\theta)d\phi]^{2} , \nonumber 
\end{eqnarray}

\noindent which yields the de Sitter space $S^{3}$. 
The radius of the $3$-sphere is scaled by the same factor: 
$R=(1/\nu)\tilde{R}=(1/\nu)2$. The parameters $\theta$, $\phi$, 
$\psi$ respectively represent the half-length of arcs which 
corrrespond to the Eulerian angles $\tilde{\theta}=\nu\theta$, 
$\tilde{\phi}=\nu\phi$, $\tilde{\psi}=\nu\psi$ on the $3$-sphere 
of radius $R=2/\nu$. Thus (\ref{metricS3}) is the metric 
on the $3$-sphere $S^{3}$ of radius $R=2/\nu$ which is 
parameterized in terms of the (half) Eulerian arclengths. The 
co-frame (\ref{coframe}) determines a unique orientation on 
this sphere. The scalar curvature $\mathcal{R}$ of this $3$-sphere 
is determined by its radius $R$ 

\begin{eqnarray} \label{cosmologicalconstant}
{\mathcal{R}} =\frac{6}{R^{2}} = \frac{3}{2} \,\, \nu^{2} ,
\end{eqnarray}

\noindent \cite{HouHou}. Note that the arclength parameters
are independent of the length scale determined by the inverse
topological mass. We shall only consider rotational degrees
of freedom which are associated with the intrinsic arclengths.

   This $3$-sphere $S^{3}$ can be embedded into the euclidean space 
$\mathbf{R^{4}}$

\begin{eqnarray} \label{embeddingS3}
(y^{1})^{2}+(y^{2})^{2}+(y^{3})^{2}+(y^{4})^{2}=R^{2} 
\hspace*{3mm} ,  \hspace*{3mm}
R=\frac{2}{\nu}, 
\end{eqnarray}

\noindent \cite{ANS} by the correspondence

\begin{eqnarray} \label{correspondenceS3}
& & y^{1}= R \cos \left ( \nu\frac{\theta}{2} \right ) 
\cos \left ( \nu\frac{\psi+\phi}{2} \right ) 
\hspace*{3mm} , \hspace*{3mm}
y^{2}= R \cos \left ( \nu\frac{\theta}{2} \right ) 
\sin \left ( \nu\frac{\psi+\phi}{2} \right ) \nonumber \\
\\
& & y^{3}= R \sin \left ( \nu\frac{\theta}{2} \right ) 
\cos \left ( \nu\frac{\psi-\phi}{2} \right ) 
\hspace*{3mm} , \hspace*{3mm}
y^{4}= R \sin \left ( \nu\frac{\theta}{2} \right ) 
\sin \left ( \nu\frac{\psi-\phi}{2} \right ) , \nonumber 
\end{eqnarray}

\noindent where $R=2/\nu$. The flat metric 

\begin{eqnarray} \label{flatR4metric}
ds^{2}= (dy^{1})^{2}+(dy^{2})^{2}+(dy^{3})^{2}+(dy^{4})^{2} ,
\end{eqnarray}

\noindent on $\mathbf{R^{4}}$ reduces to (\ref{metricS3}) 
with this correspondence. 

   The Hodge-duality relations and the Maurer-Cartan equations 
for the basis (\ref{coframe}) immediately lead to the result 
that the MCS field equation: $d(*F-\nu A)=0$ will be identically 
satisfied for the gauge potential $1$-form: $A=-(\nu/g)\omega^{3}$ 
\cite{ANS}. The field equation is given by the derivative of the 
self-duality condition: $*F-\nu A=0$ for the topologically massive 
abelian gauge fields \cite{TPN}, \cite{Deser1}. The MCS action vanishes 
for this potential: $S_{MCS}[A]=0$. Note that a change in the orientation 
leads to a change of sign in the field equation. 

  We shall define the Hopf map $S^{3} \longrightarrow S^{2}$ 
including the topological mass in section 3.2. We also scale 
the radius of the unit $2$-sphere by the same factor $1/\nu$

\begin{eqnarray} \label{embeddingS2}
(x^{1})^{2}+(x^{2})^{2}+(x^{3})^{2}=r^{2} 
\hspace*{3mm} ,  \hspace*{3mm}
r=\frac{1}{\nu} . 
\end{eqnarray}

\noindent The correspondence with the metric on $S^{2}$ 
is given by

\begin{eqnarray} \label{correspondenceS2}
x^{1}= r \sin (\nu\theta) \cos (\nu\phi) 
\hspace*{3mm} , \hspace*{3mm}
x^{2}= r \sin (\nu\theta) \sin (\nu\phi) 
\hspace*{3mm} , \hspace*{3mm}
x^{3}= r \cos (\nu\theta) ,  
\end{eqnarray}

\noindent where $r=1/\nu$. This provides an embedding of 
this $2$-sphere $S^{2}$ into the space $\mathbf{R^{3}}$ 
with radius $r=1/\nu$. The flat metric  

\begin{eqnarray} \label{flatR3metric}
ds^{2}=(dx^{1})^{2}+(dx^{2})^{2}+(dx^{3})^{2} ,
\end{eqnarray}

\noindent on $\mathbf{R^{3}}$ reduces to the metric

\begin{eqnarray} \label{defectedS2metric}
ds^{2}=d\theta^{2}+\sin^{2} \left ( \nu\theta \right ) d\phi^{2} ,
\end{eqnarray}

\noindent on $S^{2}$ with this correspondence. Thus (\ref{defectedS2metric}) 
is the metric on the $2$-sphere $S^{2}$ of radius $r=1/\nu$ which is 
parameterized in terms of the length of spherical arcs corresponding to 
the spherical angles $\tilde{\theta}=\nu\theta$, $\tilde{\phi}=\nu\phi$. 
More precisely, $\theta$ is the length of the geodesic arc from the north 
pole to the south and $\phi$ is the length of the geodesic arc on the equator. 

  The range of the parameters $\theta$, $\phi$, $\psi$ are determined by 
the topological mass. The arclength $\phi$ on the equator of $S^{2}$ uniquely 
defines the angle $\tilde{\phi}=\nu\phi$ for any parallel. The length of an 
arc, on the parallel determined by $\theta=(1/\nu)\tilde{\theta}$, which 
corresponds to the angle $\tilde{\phi}$ is

\begin{eqnarray}
\frac{\sin (\nu\theta)}{\nu} \tilde{\phi}=\sin (\nu\theta)\phi .
\end{eqnarray}

\noindent This reduces to the arclength on the equator: 
$\phi=(1/\nu)\tilde{\phi}$ when 
$\tilde{\theta}=\nu\theta=\pi/2$. The coordinate $\phi$ 
which is well-defined for any point except the poles on 
$S^{2}$ changes from $\phi=0$ to $\phi=2\pi/\nu$ 
round about any parallel. 

\subsection{The Topologically Massive Non-abelian Solution}

   First consider the solution which is given by the gauge potential 
$1$-form:

\begin{eqnarray} \label{abelianpotentialS3}
A = -\frac{\nu}{g} \, \omega^{3} \, \tau_{3}
=  -\frac{\nu}{g} \, \Big[ d\psi+ \cos(\nu\theta) d\phi \Big] \tau_{3} ,
\end{eqnarray}
   
\noindent \cite{ANS}. Here $\tau_{i}=\sigma_{i}/2i$ where $\sigma_{i}$ are 
the Pauli spin matrices. The factor $\nu/g$ yields $(1/4\pi) ng$ for the 
strength of the potential $A$ upon quantization of the topological mass 
(\ref{quantizationoftopologicalmass}). This is analogous to Dirac's 
quantization condition $eg=n$ which leads to strength $g$ for the 
non-abelian solution upon a gauge transformation of the Dirac's 
monopole potential \cite{WuYang2}, \cite{Ryder}. 

   A gauge transformation of the potential (\ref{abelianpotentialS3}) 
necessarily contains the topological mass inherently in the gauge function, 
because it also appears in the potential (\ref{abelianpotentialS3}). This 
is intuitively analogous to the gauge transformation in the non-massive
Wu-Yang monopole that contains the dimensionless factor $eg$. In our case 
this is basically due to the arclength parameterization. The gauge function 
which embeds the topologically massive abelian solution 
(\ref{abelianpotentialS3}) into the YMCS theory is an element of the group 
$SU(2)$ \cite{BhagVen}, \cite{Gottfried}. This is given by normalizing the 
radius $R$ of $S^{3}$ which is parameterized with the Eulerian arclengths as

\begin{eqnarray} \label{nonabeliangaugetransformation}
\hspace*{8mm}
 U=\frac{1}{R} \left( \begin{array}{cc}
z^{1} & z^{2} \\
-\bar{z}^{2} & \bar{z}^{1} 
\end{array} \right)  =\exp{(-\nu\gamma\tau_{3})}
\exp{(-\nu\beta\tau_{2})}
\exp{(-\nu\alpha\tau_{3})} . 
\end{eqnarray}

\noindent Here  $z^{1}=y^{1}+iy^{2}$, $z^{2}=y^{3}+iy^{4}$.
We identify the Euler parameters as 
$\alpha=\phi , \,\, \beta=\theta , \,\, \gamma=\psi$.
This is the analog of the gauge transformation which 
embeds the Dirac monopole solution into the $SU(2)$ YM 
theory \cite{WuYang2}, \cite{Ryder}. Note that 
$U \rightarrow 1$ as $\nu \rightarrow 0$. 
 
   The gauge transformation (\ref{gaugetransformation}) with 
the gauge function (\ref{nonabeliangaugetransformation}) have 
winding number $w=1$ (\ref{windingnumber}). It yields the 
potential $1$-form

\begin{eqnarray} \label{nonabelianpotential1}
& & A' = U^{-1}AU - \frac{1}{g}U^{-1}dU \\
& & \hspace*{6mm} = - \frac{\nu}{g} \,
\bigg\{ \Big[ \sin(\nu\phi)\tau_{1} - \cos(\nu\phi)\tau_{2} \Big] 
d\theta \nonumber \\
& & \hspace*{10mm}+ \Big[ \cos(\nu\theta)\cos(\nu\phi)\tau_{1}
+ \cos(\nu\theta)\sin(\nu\phi)\tau_{2} 
- \sin(\nu\theta)\tau_{3} \Big] \sin(\nu\theta) d\phi \bigg\} . \nonumber
\end{eqnarray}

\noindent This gives rise to the field $2$-form

\begin{eqnarray} \label{nonabelianfield12}
& & F' = dA' - g A' \wedge A' = U^{-1}FU \\ 
& & \hspace*{6mm} = \frac{\nu^{2}}{g}
\Big[ \sin(\nu\theta)\cos(\nu\phi)\tau_{1}
+ \sin(\nu\theta)\sin(\nu\phi)\tau_{2} 
+\cos(\nu\theta)\tau_{3} \Big] \nonumber \\
& & \hspace*{83mm} \sin(\nu\theta)  d\theta\wedge d\phi . \nonumber
\end{eqnarray}

\noindent The dual field $1$-form covariantly transforms as: 
$*F'=U^{-1}*FU$.  The field equation (\ref{YMCSfieldequation}) 
and the Bianchi identity (\ref{Bianchi}) are identically 
satisfied since they also covariantly transform. This is 
the topologically massive \textit{non-abelian} Wu-Yang 
solution which is the generalization of the non-massive 
Wu-Yang solution given in \cite{WuYang2}. 

    We have three observations which are based on \textit{dimensional}
arguments in the equations (\ref{abelianpotentialS3}), 
(\ref{nonabeliangaugetransformation}), (\ref{nonabelianpotential1}), 
(\ref{nonabelianfield12}). \\

\noindent (i) The gauge function necessarily contains the 
topological mass as in (\ref{nonabeliangaugetransformation}). \\
(ii) The strength $\nu/g$ of the abelian gauge potential 
(\ref{abelianpotentialS3}) is crucial for finding \\
\hspace*{6mm} the field $2$-form (\ref{nonabelianfield12}) in the non-abelian 
case. One finds the correct ex\-\hspace*{7mm}pression for the field $2$-form 
with this choice. \\
(iii) This is associated with the quantization of the 
topological mass.\\

   Thus if the strength of the potential is given by $\nu/g$ then we 
inevitably arrive at condition (\ref{quantizationoftopologicalmass}), 
because this yields $\nu/g=(1/4\pi) ng$ as the strength of the potential. 
Conversely, if one starts with a potential with strength $(1/4\pi)ng$, 
then one needs to use (\ref{quantizationoftopologicalmass}) in finding 
the non-abelian gauge potential $A'$ (\ref{nonabelianpotential1}) 
and field $F'$ (\ref{nonabelianfield12}). Moreover the YMCS field equation 
reduces to condition (\ref{quantizationoftopologicalmass}) which will be 
identically satisfied. Note that in this discussion, number $n$ is a free 
parameter. The gauge coupling strength and the electric charge are denoted 
with $g$. The factor of $1/4\pi$ in the topological mass 
(\ref{quantizationoftopologicalmass}) is included in the strength of the 
potential for the sake of conciseness.

   Action (\ref{YMCSaction}) for the potential 
(\ref{nonabelianpotential1}) reduces to the Yang-Mills 
term because the Chern-Simons piece vanishes 

\begin{eqnarray}
S_{YMCS}[ A'] = -8\pi^{2}\frac{\nu}{g^{2}} = -2\pi n .
\end{eqnarray}

\noindent This is equal to change $W$ (\ref{Windingnumber}) in the action 
due to the gauge transformation (\ref{nonabeliangaugetransformation}) which 
have winding number $w=1$. The other term which arise in the action as a 
result of the gauge transformation (\ref{nonabeliangaugetransformation}) 
vanishes. The quantization of the topological mass also leads to 

\begin{eqnarray} \label{ScalaRg}
{\mathcal{R}} \sim n^{2}g^{4} ,
\end{eqnarray}

\noindent for the scalar curvature (\ref{cosmologicalconstant}).

\section{The Abelian Gauge Theory}

\subsection{Maxwell-Chern-Simons Theory}

   The action for MCS theory is

\begin{eqnarray} \label{MCSaction}
S_{MCS} &=& S_{M}+S_{CS} \\
&=& -\frac{1}{2} \,\, \frac{1}{2\pi} \, \left( \int F \wedge *F  
- \nu \int F \wedge A \right) . \nonumber 
\end{eqnarray}

\noindent For later convenience, we explicitely write $1/2\pi$ in 
(\ref{quantizationoftopologicalmass}) as an overall factor in the 
action. We also adopt a slight change of convention in the abelian 
potential: $A=-(1/2)(\nu/g) \omega^{3}$ and the topological mass is 
now given as: $\nu = ng^{2}$. The MCS action (\ref{MCSaction}) yields 
the field equation

\begin{eqnarray} \label{MCSfieldequation}
\left ( *d - \nu \right ) *F = 0.
\end{eqnarray}

\noindent The field $2$-form also satisfies the Bianchi 
identity $dF=0$. We can easily derive

\begin{eqnarray} \label{InhomLaplace}
( \triangle - \nu^{2} ) F =0 \hspace*{5mm} , \hspace*{5mm} 
( \triangle - \nu^{2} ) *F = * ( \triangle - \nu^{2} ) F = 0 ,
\end{eqnarray}

\noindent where $\triangle = ( d + \delta )^{2}$ is the 
Laplace-Beltrami operator. The operator $\triangle - \nu^{2}$ 
is factorizable as 

\begin{eqnarray} \label{FactInhomLaplace}
\triangle - \nu^{2} = ( d* - \nu )(d* + \nu) 
= ( d* + \nu )(d* - \nu) . 
\end{eqnarray}

   We find 

\begin{eqnarray} \label{MCSpotentialequation}
\left ( *d -\nu \right ) A = \nu \alpha ,
\end{eqnarray}

\noindent if we integrate (\ref{MCSfieldequation}), after taking the 
Hodge-dual. Here $\alpha$ is a closed $1$-form: $d\alpha =0$ which can 
be made to vanish by means of a gauge transformation if locally 
$\alpha = dB$. Thus the topologically massive field locally determines 
the potential up to a gauge term via the self-duality equation

\begin{eqnarray} \label{potentialfromfield}
\left( *d-\nu \right) A' =0, \hspace*{5mm} \hspace*{5mm} F'=F=dA' ,
\end{eqnarray}

\noindent where $ A' = A +  \alpha$ \cite{TPN}, \cite{Deser1}. 
Therefore the field equation is identically satisfied for a self-dual 
field/potential $A'$. Because the equations (\ref{MCSfieldequation}) 
and (\ref{potentialfromfield}) are symmetric under the interchange 
$*F \leftrightarrow \nu A$. Hence we can also treat the dual-field 
$1$-form  $*F$ as a gauge potential in the field equation 
(\ref{MCSfieldequation}). As a result the field equation 
(\ref{MCSfieldequation}) reduces to finding the strength of the potential 
which is determined by the field strength itself

\begin{eqnarray} \label{Selfdualfieldequation}
F'=dA'=\frac{1}{\nu} d*F=F .
\end{eqnarray}

\noindent The Bianchi identity $dF'=0$ is also satisfied. The Maxwell 
and the Chern-Simons terms in the MCS action (\ref{MCSaction}) 
interchange under this symmetry as $F'=F$ is kept fixed. 

  We can define yet another potential $\tilde{A}$ by the 
transformation

\begin{eqnarray} \label{generalizedconnectionk=1}
\tilde{A}=\frac{1}{\nu} \, \left( *d+\nu \right) A = A + \frac{1}{\nu} *F ,
\end{eqnarray}

\noindent which is motivated by this interchange symmety. The right-hand 
side in (\ref{generalizedconnectionk=1}) is the self-duality equation with 
an opposite sign for the topological mass. The new potential $\tilde{A}$ 
transforms as a connection under the abelian gauge transformations. The 
potential $\tilde{A}$ and the field 

\begin{eqnarray} \label{generalizedfield2formk=1}
\tilde{F} = d\tilde{A} = \frac{1}{\nu} \left ( d* + \nu \right ) F ,
\end{eqnarray}

\noindent respectively satisfy the equations (\ref{MCSpotentialequation}) 
and (\ref{MCSfieldequation}) by virtue of (\ref{InhomLaplace}), 
(\ref{FactInhomLaplace}). The field $\tilde{F}$ also satisfies the 
Bianchi identity $d\tilde{F}=0$. The equations (\ref{MCSfieldequation}) 
and (\ref{MCSpotentialequation}) in turn yield $\tilde{A} = 2A + \alpha$ 
and $\tilde{F}= 2F$. A variant of this transformation 
(\ref{generalizedconnectionk=1}) is also observed in \cite{Itzhaki} 
and similar features are used in \cite{Deser2, Deser3}. In fact, with 
a similar reasoning, one can introduce higher order terms of type 
$(\frac{1}{\nu}*d)^{i} A$ where $i=1, 2, 3, ...$. The topologically massive 
gauge theory is related to the CS theory through such an expansion of field 
redefinition in \cite{GL}, \cite{LJSSVV, LJSVV}.

   The equation (\ref{generalizedfield2formk=1}) leads to the trivial 
conclusion that the field $2$-form $F$ which satisfies the field equation 
(\ref{MCSfieldequation}) becomes a source for the field equation with an 
opposite sign for the topological mass

\begin{eqnarray} \label{MCSfieldequationRevtopmass}
\left ( *d + \nu \right ) *F = J .
\end{eqnarray}
 
\noindent Here the source $1$-form is given as $J=2\nu *F$. In this 
case: $d*J=0$ since $dF=0$. The self-duality $*F=\nu A$ yields: 
$J=2\nu^{2} A= 2\triangle A$ (\ref{InhomLaplace}). 

   We shall see in section 3.4 that we can interpret 
(\ref{generalizedconnectionk=1}) as a gauge transformation  

\begin{eqnarray} \label{abeliangaugetransformation}
A'=A - \frac{i}{g}U^{-1}dU ,
\end{eqnarray}

\noindent of the potential $A$ on the $2$-sphere. We find that 
the gauge function is 

\begin{eqnarray} \label{generalizedgaugetransformation}
U = \exp \left ( i \frac{g}{\nu} \oint *F \right ) 
= \exp \left[ ig \oint ( A + \alpha ) \right] ,
\end{eqnarray}

\noindent upon the identification

\begin{eqnarray}
\frac{1}{\nu} \, *F=-\frac{i}{g} d \ln U ,
\end{eqnarray}

\noindent and using the self-duality equation 
(\ref{potentialfromfield}). This is the gauge 
function which will arise in the topologically 
massive Wu-Yang solution in terms of a loop 
integral that yields the electric charge.

\subsection{The Hopf Map with the Topological Mass}

   In this section we present the Hopf map $S^{3} \longrightarrow S^{2}$ 
including the topological mass. This leads to local topologically massive 
Wu-Yang potentials on $S^{2}$ upon using specific sections of $S^{3}$ which 
is locally given as a $S^{1}$ bundle over $S^{2}$: 
$S^{3} \approx S^{1} \times S^{2}$ \cite{M,R,T}. The difference from 
the non-massive Wu-Yang monopole \cite{WuYang2}, \cite{Ryder} which is 
essentially described by a $U(1)$ principal bundle over the unit $2$-sphere 
is the introduction of a natural scale of length by the topological mass. 
We refer the reader to \cite{Pen, Berger, Thurston, AzcarragaIzquierdo, 
Vassiliev, Spivak, Steenrod, BottTu, GreuHalpVan} for the basic concepts 
of the Hopf map and its relation to Dirac monopole \cite{M, R, T} which we 
adapt here with the inclusion of the topological mass.  For a discussion of 
Wu-Yang type multi-monopole solutions see \cite{Popov}.

   The Hopf map $S^{3} \longrightarrow S^{2}$ is given as

\begin{eqnarray} \label{Hopfmap}
& & x^{1} = 2\frac{r}{R^{2}} \left ( y^{1}y^{3}+y^{2}y^{4} \right ) ,
\nonumber \\
& & x^{2} = 2\frac{r}{R^{2}} \left ( y^{2}y^{3}-y^{1}y^{4} \right ) , \\
& & x^{3} = \frac{r}{R^{2}} \left [ (y^{1})^{2}+ (y^{2})^{2}
-(y^{3})^{2}-(y^{4})^{2} \right ] , \nonumber 
\end{eqnarray}

\noindent \cite{M}, \cite{R,T}. The sections of the $3$-sphere 
(\ref{correspondenceS3}) corresponding to $\phi=-\psi$ and $\phi=\psi$ 
are respectively 

\begin{eqnarray} \label{sectionnorth}
& & z^{1} =R \cos \left( \frac{\nu\theta}{2} \right)
= R \frac{r}{\sqrt{r^{2}+|w|^{2}}} , \\
& & z^{2} = R \sin \left( \frac{\nu\theta}{2} \right) \exp ( -i\nu\phi ) 
= R \frac{w}{\sqrt{r^{2}+|w|^{2}}} , \nonumber \\
& & \hspace*{10mm}
w=r\frac{x^{1}-ix^{2}}{r+x^{3}} = r\frac{z^{2}}{z^{1}}
=r\tan \left( \frac{\nu\theta}{2} \right) \exp ( -i\nu\phi ) ,\nonumber
\end{eqnarray}

\noindent and

\begin{eqnarray} \label{sectionsouth}
& & z^{1} =R \cos \left( \frac{\nu\theta}{2} \right) \exp ( i\nu\phi ) 
= R \frac{z}{\sqrt{r^{2}+|z|^{2}}} , \nonumber \\
& & z^{2} =R \sin \left( \frac{\nu\theta}{2} \right)
= R \frac{r}{\sqrt{r^{2}+|z|^{2}}} , \\
& & \hspace*{10mm}
z=r\frac{x^{1}+ix^{2}}{r-x^{3}} = r\frac{z^{1}}{z^{2}}
=r\cot \left( \frac{\nu\theta}{2} \right) \exp ( i\nu\phi ) . \nonumber 
\end{eqnarray}

\noindent Here $z$ and $w$ are respectively the stereographic projection 
coordinates of the southern hemi-sphere $U_{S}$ and the northern hemi-sphere 
$U_{N}$ of $S^{2}$ projected from the north pole and the south pole to the 
equatorial plane. They satisfy $zw=r^{2}$ (\ref{sectionnorth}), 
(\ref{sectionsouth}). 

  The inverse image of a point in $S^{2}$ is given by either: 
$z^{1}=\frac{1}{r} zz^{2}$ or $z^{2}=\frac{1}{r} wz^{1}$ where 
$| z^{1} |^{2}+| z^{2} | ^{2}=R^{2}$. These are the intersections 
of $S^{3}$ with hyperplanes passing through the origin. Thus for 
every point in $S^{2}$, there exists a $S^{1}$ fibre which is a great 
circle of radius $R=2/\nu$ in $S^{3}$ \cite{Thurston, AzcarragaIzquierdo, 
Vassiliev, Spivak}.  These are parameterized as $\exp ( -\nu\psi\tau_{3} )$ 
or simply $\exp ( i\nu\psi/2 )$ \cite{AzcarragaIzquierdo}. We shall present 
a stereographic view of these fibres in section 4. 

  The $3$-sphere $S^{3}$ is locally given as $S^{1} \times S^{2}$.
However $S^{1} \times S^{2}$ is globally distinct \cite{R}. Because on 
$S^{2}$ the angular coordinate $\tilde{\phi}=\nu\phi$ is undefined for 
$\tilde{\theta}=\nu\theta =0$ and $\tilde{\theta}=\nu\theta =\pi$. 
Nevertheless, the two local sections given in the equations 
(\ref{sectionnorth}), (\ref{sectionsouth}) are respectively well-defined 
when $\tilde{\theta}=\nu\theta=0$ and $\tilde{\theta}=\nu\theta=\pi$ that 
is on $U_{N}$ and $U_{S}$. The transition function on the equator 
$U_{N} \bigcap U_{S}$ is given by $\exp ( -2\nu\phi\tau_{3} )$ that is 
$\exp(i\nu\phi)$ as one can see from equations (\ref{sectionnorth}), 
(\ref{sectionsouth}) in analogy with the non-massive case 
\cite{AzcarragaIzquierdo}. Note that the gauge group $U(1)$ is given 
by normalizing the radius of $S^{1}$. 

   The $1$-form $(1/2)\omega^{3}$ (\ref{coframe}) defines a connection 
on $S^{3}$ which is considered as a $S^{1}$ bundle over $S^{2}$ analogous 
to the non-massive case. This gives rise to gauge potential
$A=-(1/2)(\nu/g)\omega^{3}$. The strength $-(1/2)(\nu/g)$ of the potential 
reduces to $-(1/2)ng$ upon adopting the quantization of the topological 
mass: $\nu=ng^{2}$. The abelian potential

\begin{eqnarray} \label{AbelianpotentialS3}
& & A =  - \frac{1}{2} \, \frac{\nu}{g} \, 
\Big[ d\psi+ \cos(\nu\theta) d\phi \Big] \\ 
& & \hspace{4mm} = - \frac{1}{g} \, \frac{1}{R^{2}} \, 
\left ( -y^{2}dy^{1}+y^{1}dy^{2}-y^{4}dy^{3}+y^{3}dy^{4} \right ) , 
\nonumber
\end{eqnarray}

\noindent is globally defined on $S^{3}$. This yields the field $2$-form

\begin{eqnarray} \label{fieldstrengthS3}
& & F= \frac{1}{2} \, \frac{\nu^{2}}{g} \, 
\sin ( \nu\theta ) d\theta \wedge d\phi \\
& & \hspace*{4mm} = - \frac{1}{g} \, \frac{2}{R^{2}} \,
\left ( dy^{1} \wedge dy^{2} + dy^{3} \wedge dy^{4} \right ) ,
\nonumber
\end{eqnarray}

\noindent which is closed and also exact on $S^{3}$. Because all closed 
forms on $3$-sphere are exact \cite{M,R,T}. The potential 
(\ref{AbelianpotentialS3}) satisfy the self-duality equation 
(\ref{potentialfromfield}).

\begin{figure}[!ht]
\begin{center}
\includegraphics[scale=0.30]{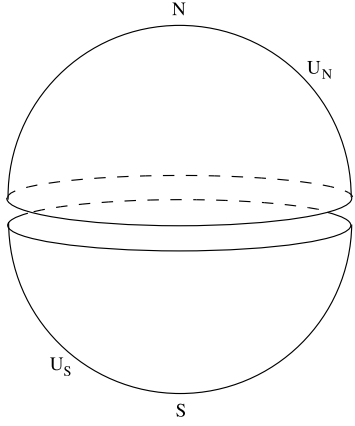}
\end{center}
\caption{$A^{N}$ and $A^{S}$ are respectively well-defined over 
$U_{N}$ and $U_{S}$.}  
\end{figure}

  However the field $2$-form is not exact although it is closed on the 
$2$-sphere $S^{2}$. If it were exact it would yield no magnetic charge 
\cite{ANS} upon using the Stokes theorem. In other words, there exists 
no globally defined potential $A$ on $S^{2}$ such that $F=dA$. Therefore 
the potential $1$-forms 

\begin{eqnarray} \label{AbelianpotentialS2}
& & A^{N} = + \, \frac{1}{2} \, \frac{\nu}{g} \, 
\Big[ 1 - \cos(\nu\theta) \Big] d\phi
=\frac{1}{2} \, \frac{1}{g} \, \, 
\frac{-x^{2}dx^{1}+x^{1}dx^{2}}{r(r+x^{3})} , \nonumber \\
\\
& & A^{S} = - \, \frac{1}{2} \, \frac{\nu}{g} \, 
\Big[ 1 + \cos(\nu\theta) \Big] d\phi
=-\frac{1}{2} \, \frac{1}{g} \, \, 
\frac{-x^{2}dx^{1}+x^{1}dx^{2}}{r(r-x^{3})} , \nonumber 
\end{eqnarray}

\noindent on $S^{2}$ are locally given by projections of the specific sections 
$\phi=-\psi$ and $\phi=\psi$ (\ref{sectionnorth}), (\ref{sectionsouth}) of $A$ 
(\ref{AbelianpotentialS3}) on $S^{3}$ onto $S^{2}$ using the Hopf map. These 
are respectively well-defined on the northern and the southern hemispheres: 
$U_{N}$ and $U_{S}$ as shown in Figure 1. The potential $1$-forms 
(\ref{AbelianpotentialS2}) lead to topologically massive Wu-Yang 
potentials 

\begin{eqnarray} \label{AbelianWuYangpotentialS2}
& & A^{N} =+ \, \frac{1}{2} \, \frac{\nu}{g} \, 
\frac{\Big[ 1 - \cos(\nu\theta) \Big]}{\sin(\nu\theta)} \, e^{(\phi)} ,
\nonumber \\
& & \hspace*{60mm} e^{(\phi)}=\sin(\nu\theta) d\phi , \\
& & A^{S} =- \, \frac{1}{2} \, \frac{\nu}{g} \, 
\frac{\Big[ 1 + \cos(\nu\theta) \Big]}{\sin(\nu\theta)} \, e^{(\phi)} , 
\nonumber 
\end{eqnarray}

\noindent in the \textit{orthonormal frame} on $S^{2}$. The potentials 
$A^{N}$ and $A^{S}$ have singularities at $\theta=\frac{\pi}{\nu}$ and 
$\theta=0$ which respectively correspond to the negative and the positive 
$x^{3}$- axis. These yield the same field $2$-form

\begin{eqnarray} \label{abelianfield2form}
& & F^{N}=F^{S}= F 
=\frac{1}{2} \, \frac{\nu^{2}}{g} \, \sin (\nu\theta) d\theta \wedge d\phi \\
& & \hspace*{27mm} 
= \frac{1}{2} \, \frac{1}{g} \, \frac{1}{r^{3}} \, \left ( x^{1}dx^{2} 
\wedge dx^{3} + x^{2}dx^{3} \wedge dx^{1} 
+ x^{3}dx^{1} \wedge dx^{2} \right ) , \nonumber
\end{eqnarray}   

\noindent on $S^{2}$. 

  We could have considered the sections N: $\psi = -\phi$ and S: 
$\psi = \phi$ for the potential (\ref{abelianpotentialS3}) and the 
gauge function (\ref{nonabeliangaugetransformation}) in embedding the 
abelian solution into the YMCS theory in section 2.3. It is straightforward 
to check that the equations (\ref{nonabelianpotential1}), 
(\ref{nonabelianfield12}) are satisfied for these sections. The stringy 
singularities of the potentials $A^{N}$, $A^{S}$ (\ref{AbelianpotentialS2}) 
vanish after these gauge transformations.

  A potential as a local expression for a connection in a bundle is 
determined by a choice of a local section in this bundle \cite{Nakahara}. 
In our solution the term $\alpha$ (\ref{potentialfromfield}) is associated 
with the specific local sections in the Hopf map. The extra $\alpha$ terms 
for potentials $A^{N}$ and $A^{S}$ (\ref{AbelianpotentialS2}) on $S^{2}$ 
vanish upon a consistent choice of the appropriate local sections N: 
$\phi=-\psi$ and S: $\phi=\psi$ for  dual-field $1$-forms: $*F^{N}=(*F)^{N}$, 
$*F^{S}=(*F)^{S}$ while projecting $S^{3}$ onto $S^{2}$. Otherwise one needs 
to use abelian gauge transformations on $S^{3}$ in order to make these terms 
vanish. The MCS field equation (\ref{MCSfieldequation})

\begin{eqnarray} \label{MCSfieldequationS2}
d*F^{N} - \nu F^{N} = 0 \hspace*{5mm} , \hspace*{5mm}
d*F^{S} - \nu F^{S} = 0 ,
\end{eqnarray}

\noindent and also the self-duality condition (\ref{potentialfromfield})

\begin{eqnarray} \label{potentialfromfieldS2}
*F^{N} - \nu A^{N} =0 \hspace*{5mm} , \hspace*{5mm} 
*F^{S} - \nu A^{S} =0 ,
\end{eqnarray}

\noindent are satisfied on each local chart $U_{N}$ and $U_{S}$ of 
$S^{2}$ for the appropriate local potentials $A^{N}$ and $A^{S}$ 
(\ref{AbelianpotentialS2}). Thus the field determines the local gauge 
potential on the $2$-sphere up to a gauge term. The Wu-Yang solution
that is based on seaming these solutions along the equator 
$U_{N} \bigcap U_{S}$ by means of a gauge transformation exhibits 
this gauge term. 

   The Hodge-duality is based on a locally consistent choice of a global 
orientation which is determined by the co-frame (\ref{coframe}) on $S^{3}$. 
This orientation is locally equivalent to the ``product orientation'' which 
is induced by the orientations of $S^{1}$ and $S^{2}$, \cite{Vassiliev}, 
\cite{Steenrod, BottTu, GreuHalpVan}. The Wu-Yang solution yields a locally 
consistent global orientation. 

   We present three special cases of interest concerning the transformation 
of coordinates in the Hopf map (\ref{Hopfmap}). In the first case, the 
transformation 

\begin{eqnarray} \label{Trf1}
y^{1} \rightarrow -y^{4} \hspace*{5mm} , \hspace*{5mm} 
y^{2} \rightarrow -y^{3} \hspace*{5mm} , \hspace*{5mm} 
y^{3} \rightarrow y^{2} \hspace*{5mm} , \hspace*{5mm} 
y^{4} \rightarrow y^{1} ,
\end{eqnarray}

\noindent which is equivalently given by 
$\tilde{\theta} \rightarrow \pi - \tilde{\theta}$, 
$\tilde{\phi} \rightarrow \pi + \tilde{\phi}$, 
$\tilde{\psi} \rightarrow 2\pi - \tilde{\psi}$ 
on $S^{3}$ (\ref{correspondenceS3}) induces 
the anti-podal map

\begin{eqnarray} \label{Trf1S2} 
x^{1} \rightarrow -x^{1} \hspace*{5mm} , \hspace*{5mm} 
x^{2} \rightarrow -x^{2} \hspace*{5mm} , \hspace*{5mm} 
x^{3} \rightarrow -x^{3} ,
\end{eqnarray}

\noindent on $S^{2}$ (\ref{correspondenceS2}), (\ref{Hopfmap}). 
This leads to $A \rightarrow -A$ (\ref{AbelianpotentialS3}), 
$F \rightarrow -F$ (\ref{fieldstrengthS3}) and $A^{N} \rightarrow -A^{S}$, 
$A^{S} \rightarrow -A^{N}$ (\ref{AbelianpotentialS2}), 
$F^{N/S} \rightarrow -F^{S/N}$ (\ref{abelianfield2form}). The minus ($-$) 
signs in the potential and the field vanish if we redefine the gauge coupling 
constant: $g \rightarrow -g$. This transformation amounts to a change of the 
special sections (\ref{sectionnorth}), (\ref{sectionsouth}) of $S^{3}$. The 
second transformation 

\begin{eqnarray} \label{Trf2} 
y^{1} \rightarrow y^{3} \hspace*{5mm} , \hspace*{5mm} 
y^{2} \rightarrow y^{4} \hspace*{5mm} , \hspace*{5mm}
y^{3} \rightarrow y^{1} \hspace*{5mm} , \hspace*{5mm} 
y^{4} \rightarrow y^{2} ,
\end{eqnarray}

\noindent leads to 

\begin{eqnarray} \label{Trf2S2} 
x^{1} \rightarrow x^{1} \hspace*{5mm} , \hspace*{5mm} 
x^{2} \rightarrow -x^{2} \hspace*{5mm} , \hspace*{5mm} 
x^{3} \rightarrow -x^{3} ,
\end{eqnarray}

\noindent on $S^{2}$. This yields $A \rightarrow A$ 
(\ref{AbelianpotentialS3}), $F \rightarrow F$ 
(\ref{fieldstrengthS3}) and $A^{N} \rightarrow A^{S}$, 
$A^{S} \rightarrow A^{N}$ (\ref{AbelianpotentialS2}), 
$F^{N/S} \rightarrow F^{S/N}$ (\ref{abelianfield2form}). 
The third case is 

\begin{eqnarray} \label{Trf3}
y^{1} \rightarrow y^{4} \hspace*{5mm} , \hspace*{5mm} 
y^{2} \rightarrow y^{3} \hspace*{5mm} , \hspace*{5mm}
y^{3} \rightarrow y^{2} \hspace*{5mm} , \hspace*{5mm} 
y^{4} \rightarrow y^{1} ,
\end{eqnarray}

\noindent which leads to
 
\begin{eqnarray} \label{Trf3S2}
x^{1} \rightarrow x^{1} \hspace*{5mm} , \hspace*{5mm} 
x^{2} \rightarrow x^{2} \hspace*{5mm} , \hspace*{5mm} 
x^{3} \rightarrow -x^{3} .
\end{eqnarray}

\noindent This yields $A \rightarrow -A$ 
(\ref{AbelianpotentialS3}), $F \rightarrow -F$ 
(\ref{fieldstrengthS3}) and $A^{N} \rightarrow -A^{S}$,
$A^{S} \rightarrow -A^{N}$ (\ref{AbelianpotentialS2}), 
$F^{N/S} \rightarrow -F^{S/N}$ (\ref{abelianfield2form}). 
The three transformations (\ref{Trf1}), (\ref{Trf2}), 
(\ref{Trf3}) are orientation preserving.

\subsection{The Topologically Massive Wu-Yang Solution }

  The abelian topologically massive Wu-Yang solution is based on patching 
up the local solutions by means of a gauge transformation which contains 
the topological mass as in the non-abelian case. The potential $1$-forms 
$A^{N}$ and $A^{S}$ (\ref{AbelianpotentialS2}) differ by an abelian gauge 
transformation 

\begin{eqnarray} \label{WuYanggaugetransformation}
A^{N}=A^{S} - \frac{i}{g}U^{-1}dU ,
\end{eqnarray}

\noindent where $U=\exp{(i\nu\phi)}$. If the strength of the potentials 
$A^{N}$ and $A^{S}$ (\ref{AbelianpotentialS2}) were $\pm(1/2)ng$ then we 
would find $\nu = ng^{2}$ in the gauge function. This conversely yields 
$\pm(1/2)\nu/g=\pm(1/2)ng$ as the strength of the potentials. The gauge 
function $U=\exp{(i\nu\phi)}$ is analogous to that of the non-massive 
Wu-Yang monopole \cite{WuYang2} which contains the dimensionless factor 
$eg$. The difference lies in the physical dimensions: 
$[ \nu ]=[g]^{2}=L^{-1}$. In this case, the phase $\nu\phi$ defines the 
dimensionless angle: $\tilde{\phi}=\nu\phi$ in terms of which the gauge 
function is single-valued: $0 \leq \tilde{\phi}=\nu\phi \leq 2\pi $.  
One could discuss quantization of the topological mass using the 
arclength in units of radius that is angle on the circle of unit 
radius $g=1$ ignoring the relevant physical \textit{dimensions}.

   The magnetic charge $Q$ \cite{ANS} is 

\begin{eqnarray} \label{charge}
Q \equiv \oint_{S^{2}} F = 2 \pi \frac{1}{g} , 
\end{eqnarray}

\noindent (\ref{abelianfield2form}) as in the non-massive case.
We decompose this integral into terms

\begin{eqnarray} \label{chargedecomposed}
Q = \int _{S^{2}} F = \oint_{U_{N}} dA^{N} +  \oint_{U_{S}} dA^{S} ,
\end{eqnarray}

\noindent which are locally well-defined over the charts $U_{N}$ and $U_{S}$. 
These reduce to loop integrals 

\begin{eqnarray} \label{chargegaugefunction}
Q= \oint_{P:\tilde{\theta}=\frac{\pi}{2}} (A^{N}-A^{S}) ,
\end{eqnarray}

\noindent of the potentials $A^{N}$, $A^{S}$ over the common 
boundary $U_{N} \bigcap U_{S}$, the equator determined by the 
parallel $P:\tilde{\theta}=\nu\theta=\frac{\pi}{2}$, if we use 
Stokes theorem. We can write this as

\begin{eqnarray} \label{gaugefunctioncharge}
Q= -\frac{i}{g}\oint_{P:\tilde{\theta}=\frac{\pi}{2}} d(\ln U) ,
\end{eqnarray}

\noindent using the equation (\ref{WuYanggaugetransformation}). 
The gauge function is $U=\exp{(igQ)}$ in terms of the magnetic 
charge (\ref{charge}). 

   The equations (\ref{charge}), (\ref{chargedecomposed}), 
(\ref{chargegaugefunction}), (\ref{gaugefunctioncharge}) are 
basically the sum of integrals of the field equations 
(\ref{MCSfieldequationS2}). Because the field strength determines 
the local potentials via the self-duality conditions 
(\ref{potentialfromfieldS2}). Therefore the gauge function can 
equivalently be given in terms of the electric charge as we noted 
in (\ref{generalizedgaugetransformation}). This construction is 
valid for any parallel $P$ which is determined by the angle 
$\tilde{\theta}=\nu\theta$ since the polar terms of $A^{N}$ and $A^{S}$ 
cancel out in (\ref{chargegaugefunction}). For $\tilde{\theta}=0$ or 
$\tilde{\theta}=\pi$ the expression (\ref{chargegaugefunction}) reduces 
to integration of one of the field equations (\ref{MCSfieldequationS2}) 
via the Stokes theorem.

\subsection{The Wu-Yang Solution and The Self-Duality}

  The local solutions (\ref{MCSfieldequationS2}), 
(\ref{potentialfromfieldS2}) have a composite structure which 
consists of both magnetic and electric charges. This is basically due to 
local character of the field equation. The Wu-Yang solution is simply based 
on patching up these local solutions by means of a gauge transformation. 
Because the field strength locally determines the potential up to a gauge 
term. The underlying reason for interpreting the equation 
(\ref{generalizedconnectionk=1}) as a gauge transformation 
(\ref{generalizedgaugetransformation}) is the field equation 
itself which leads to this composite structure. 

   The integration of either of the field equations 
(\ref{MCSfieldequationS2}) over the relevant chart 
yields 

\begin{eqnarray} \label{fluxcharge}
\Phi - \nu Q = 0 . 
\end{eqnarray}

\noindent Here $\Phi$ is the electric charge which is 
defined as 

\begin{eqnarray} \label{Electricharge}
\Phi \equiv \int_{S^{2}} d*F .
\end{eqnarray}

\noindent This reduces to 

\begin{eqnarray}  \label{ElectrichargeStokes}
\Phi = \oint_{P} *F = \nu  \oint_{P} A , 
\end{eqnarray}

\noindent a loop integral over the boundary of the relevant chart 
which is given by the parallel $P$ upon using Stokes theorem.
Here the last equality follows from the equation 
(\ref{potentialfromfieldS2}). Thus

\begin{eqnarray} \label{loopintegral}
Q= \frac{1}{\nu} \Phi = \oint_{P} A .
\end{eqnarray}

\noindent We can write the gauge function in the Wu-Yang solution as

\begin{eqnarray} \label{bothgaugefunctions}
U=\exp(igQ)=\exp \left ( i\frac{g}{\nu}\Phi \right ) ,
\end{eqnarray}

\noindent in terms of the magnetic or the electric charge. Therefore 
the two gauge functions in (\ref{generalizedgaugetransformation}) and 
(\ref{gaugefunctioncharge}) coincide. Note that the superscripts $N$/$S$ 
in the equations (\ref{fluxcharge}), (\ref{Electricharge}), 
(\ref{ElectrichargeStokes}), (\ref{loopintegral}), 
(\ref{bothgaugefunctions}) are omitted for the 
sake of conciseness.

  The magnetic and the electric charges associated with the potentials 
$A^{N}$ and $A^{S}$ on $S^{2}$ except the poles $\tilde{\theta}=\pi$ 
and $P: \tilde{\theta}=0$ are

\begin{eqnarray} \label{magelchargeNS}
& & Q^{N} =\frac{1}{\nu} \Phi^{N}= \oint_{P:\tilde{\theta}=\pi} A^{N} ,
\hspace*{14mm}  
Q^{S} = \frac{1}{\nu} \Phi^{S}= - \oint_{P:\tilde{\theta}=0} A^{S} 
\nonumber \\
& & \hspace{7mm} =2 \pi \frac{1}{g}
\hspace*{46mm}
=2 \pi \frac{1}{g} ,
\end{eqnarray}

\noindent (\ref{charge}) (\ref{Electricharge}), (\ref{ElectrichargeStokes}), 
(\ref{loopintegral}). The quantization of the topological mass: $\nu=ng^{2}$ 
leads to quantization of the electric charge: $\Phi^{N}=\Phi^{S}=2\pi ng$. 
The parallels $P: \tilde{\theta}=\pi$ and $P: \tilde{\theta}=0$ refer to very 
small loops around the poles. These equations will lead us to Archimedes map 
in section $4.2$ which yields a simple picture of the Wu-Yang construction in 
terms of area and arclength. 

\begin{figure}[!ht]
\begin{center}
\includegraphics[scale=0.65]{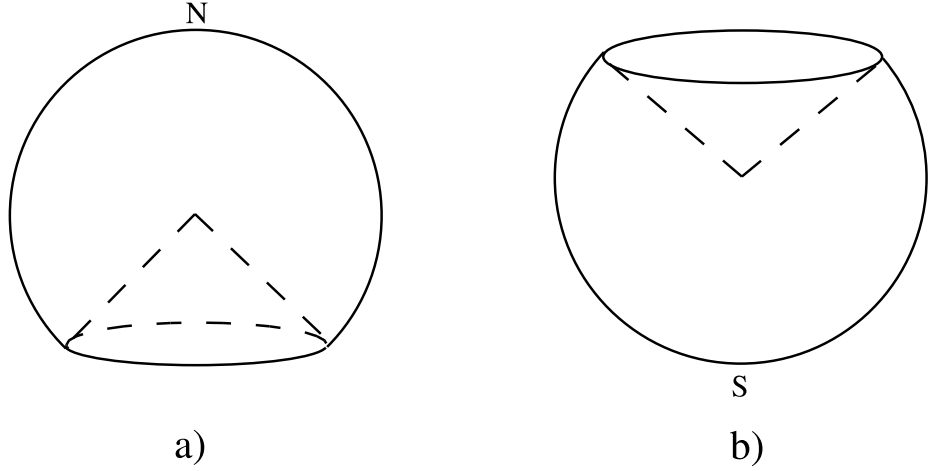}
\end{center}
\caption{The local charts of $S^{2}$ have boundaries 
at the parallels $P$ around the poles which correspond 
to singular points.}  
\end{figure}

  If the Wu-Yang construction is carried out around an arbitrary parallel 
$P: \tilde{\theta}=\nu\theta$, the magnetic and electric charges are 
\textit{partially} given by

\begin{eqnarray} \label{partialcharges}
& & Q^{N}=\frac{1}{\nu} \Phi^{N}= \oint_{P:\tilde{\theta}} A^{N} 
\hspace*{7mm} , \hspace*{7mm}
Q^{S}=\frac{1}{\nu} \Phi^{S}= - \oint_{P:\tilde{\theta}} A^{S} \nonumber \\
\nonumber \\
& & \hspace*{7mm} = \pi \frac{1}{g} \Big[ 1 - \cos(\nu\theta) \Big] 
\hspace*{7mm}  \hspace*{13mm}
= \pi \frac{1}{g} \Big[ 1 + \cos(\nu\theta) \Big].  
\end{eqnarray}

\noindent  The total magnetic and the electric charges are respectively
$Q_{T}=Q^{N}+Q^{S}$, $\Phi_{T}=\Phi^{N}+\Phi^{S}$; $Q_{T}=(1/\nu)\Phi_{T}$ 
since the two charts constitute the $2$-sphere. The gauge function is 
$U=\exp{(i\nu\phi)}$. Its line integral over the parallel yields the total 
magnetic charge $Q_{T}$ or equivalently the electric charge $\Phi_{T}$. 

   We can explicitly write the Wu-Yang construction with 
the electric charge as follows

\begin{eqnarray} \label{WuYangElectricharge}
& & \Phi = \int_{S^{2}} d*F = \int_{U_{N}} d*F^{N} + \int_{U_{S}} d*F^{S} 
\\ \nonumber
& & \hspace{4mm} =\oint_{P} (*F^{N} - *F^{S})
\\ \nonumber
& & \hspace{4mm} = \nu \oint_{P} (A^{N} - A^{S}) 
\\ \nonumber
& & \hspace{4mm} = - \frac{i}{g} \nu \oint_{P} U^{-1} dU .
\nonumber 
\end{eqnarray}

\noindent Here we used (\ref{potentialfromfieldS2}). This yields 
$U=\exp{[i(g/\nu)\Phi]}$ (\ref{bothgaugefunctions}). 

  Although the dual-field $1$-form covariantly transforms 
as: $*F'=U^{-1}*FU$ on $S^{3}$, its sections: $*F^{N}$, $*F^{S}$ on 
$S^{2}$ transform as proportional to the potentials. Because choice of 
these local sections endows us with potentials as local expressions for 
the connection which are determined by the field itself via the self-duality 
condition.

\subsection{The First Chern Number}

  Consider the generalization 

\begin{eqnarray} \label{generalizedconnectionk}
\tilde{A} = A + \frac{k}{\nu} *F ,
\end{eqnarray}

\noindent of the transformation (\ref{generalizedconnectionk=1})
where $k$ is an arbitrary constant. This is equivalent to scaling 
the potential $A$ by a dimensionless factor modulo a closed term 
(\ref{potentialfromfield}). The new potential $\tilde{A}$ transforms 
as a connection under abelian gauge transformations. The equation 
(\ref{generalizedconnectionk}) yields $\tilde{F}=d\tilde{A}=0$ 
(\ref{MCSfieldequation}) for $k=-1$,

  The sections $\tilde{A}^{N}$ and $\tilde{A}^{S}$ of the 
potential $\tilde{A}$ (\ref{generalizedconnectionk}) which 
are $1+k$ multiple of the potentials (\ref{AbelianpotentialS2}), 
(\ref{AbelianpotentialS3}) are well-defined respectively on 
$U_{N}$ and $U_{S}$. The magnetic and electric charges associated 
with the potentials $\tilde{A}^{N}$ and $\tilde{A}^{S}$ are 

\begin{eqnarray} \label{generalizedelectromagneticharge}
& & \tilde{Q} =  ( 1 + k ) Q 
\hspace*{15mm} , \hspace*{15mm}
\tilde{\Phi}= ( 1 + k ) \Phi \\
& & \hspace*{5mm} =  2 \pi \frac{1}{g} ( 1 + k )  
\hspace*{24mm} 
= 2\pi \frac{\nu}{g} (1+k) , \nonumber
\end{eqnarray}

\noindent where $\tilde{\Phi}=\nu \tilde{Q}$. This yields 
$\tilde{\Phi}=2\pi ( 1 + k ) ng$ if we use $\nu=ng^{2}$. The 
transformation (\ref{generalizedconnectionk}) changes the magnetic 
and the electric charges by $k$ units. The potentials $\tilde{A}^{N}$ 
and $\tilde{A}^{S}$ differ by a gauge transformation: 
$\tilde{U}=\exp \left [ i \left ( 1 + k \right ) \nu \phi \right ]$ 
(\ref{WuYanggaugetransformation}). The Wu-Yang construction with 
the potentials $\tilde{A}^{N}$ and $\tilde{A}^{S}$ yields 
$\tilde{U}=\exp{(ig\tilde{Q})}
=\exp{[i(g/\nu)\tilde{\Phi}]}=U^{1+k}$. 
The gauge function is single-valued in the interval 
$0 \leq \nu\phi \leq 2\pi$ if $k$ is an integer. 

  The First Chern number $C$ \cite{AzcarragaIzquierdo}, 
\cite{EguchiGilkeyHanson} is given as

\begin{eqnarray}
& & C= - \, \frac{1}{2\pi} \, g \, \tilde{Q} 
= - \, \frac{1}{2\pi} \frac{g}{\nu} \, \tilde{\Phi} \\
& &  \hspace*{5mm} 
= - ( 1 + k ) . \nonumber
\end{eqnarray}

\noindent This is determined by the winding number of the gauge function 
$\tilde{U}:S^{1} \rightarrow S^{1}$ \cite{AzcarragaIzquierdo}
as we shall discuss in section $5$.

\section{Three Geometric Structures}

   In this section we present three geometric structures which 
arise in connection with our solutions. The first is a stereographic 
view of fibers in the Hopf map. The second is Archimedes map from 
$S^{2}$ to the cylinder around it. The third is the expression of 
the geometric phase suffered by a vector upon parallel transport 
on $S^{2}$ in terms of holonomy of the topologically massive gauge 
potential or the dual-field. 

\subsection{A Stereographic View}

  In the Hopf map $S^{3} \longrightarrow S^{2}$ (\ref{Hopfmap}), for 
every point in $S^{2}$ there exists a corresponding great circle $S^{1}$ 
of radius $R=2/\nu$ in $S^{3}$. These great circles can be explicitly 
found by inverting the Hopf map. Equivalently they are given by intersections 
of $S^{3}$ with hyperplanes passing through the origin.

  The stereographic projection $S^{3} \longrightarrow \mathbf{R^{3}}$ 

\begin{eqnarray} \label{S3StereR3}
(X^{1}, X^{2}, X^{3})= \frac{R}{R-y^{1}} \left( y^{3}, y^{4}, y^{2} \right) ,
\end{eqnarray}
  
\noindent from the pole $N(R,0, 0, 0)$ of $S^{3}$ provides a stereographic 
view of these $S^{1}$ fibers, as given in Figure 3 \cite{Pen}, 
\cite{Berger, Thurston}. Each torus in the figure is made up of 
circles corresponding to points on a parallel determined by $\theta$ on 
$S^{2}$. These circles are called Villarceau circles. The circle corresponding 
to the north pole on $S^{2}$ is given by the $X^{3}$-axis. The innermost 
circle with radius $R$ in the $(X^{1}, X^{2})$ plane corresponds to the south 
pole. The tori respectively correspond to the parallels $\theta=\pi/(4\nu)$, 
$\theta=\pi/(2\nu)$ and $\theta=3\pi/(4\nu)$. The Villarceau circles are 
linked once with each other. This yields the geometric interpretation of 
the Hopf invariant in terms of the linking number of the Villarceau circles 
\cite{Thurston}, \cite{Steenrod, BottTu, GreuHalpVan}.

\begin{figure}[!hb]
\begin{center}
\includegraphics[scale=0.4]{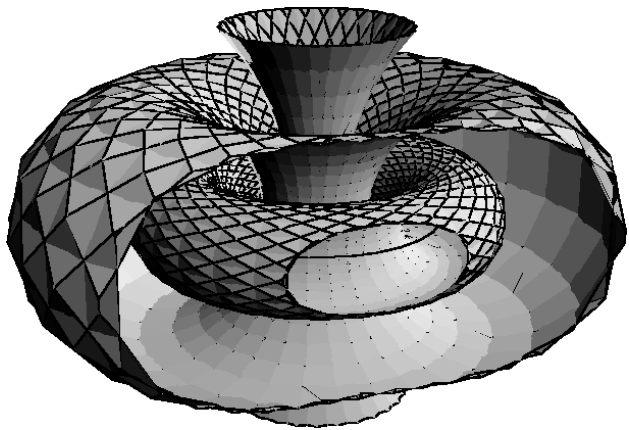}
\end{center}
\caption{The $(X^{1}, X^{2}, X^{3})$ axis are chosen in 
accordance with a right-handed coordinate system.}  
\end{figure}

\subsection{The Archimedes Map}

  The Archimedes map \cite{MdS} or the Lambert projection \cite{AF} as 
it is sometimes called from $S^{2}$ to the cylinder over its equatorial
circle provides a simple geometric explanation of the Wu-Yang construction
and the field equation in terms of area and arclength. The Archimedes map 

\begin{eqnarray}
{\mathcal{A}}(x, y, z)= \left( \frac{r}{\sqrt{r^{2}-z^{2}}} \, x, \,\, 
\frac{r}{\sqrt{r^{2}-z^{2}}} \, y, \,\, z \right) ,
\end{eqnarray}

\noindent is shown in Figure 4 
$[(x^{1}, x^{2}, x^{3}) \longrightarrow (x, y, z)]$. 
Choose a point $P$, except the poles, on $S^{2}$ and draw a straight 
line perpendicular to the axis of the cylinder which passes through 
$P$. The image of $P$ is defined as the point of intersection of this 
line with the cylinder. This yields $z=(1/\nu)\cos(\nu\theta)$ and 
$\phi$ denotes the length of arcs on parallels of the cylinder. This 
map is area preserving.

\begin{figure}[!hb]
\begin{center}
\includegraphics[scale=0.4]{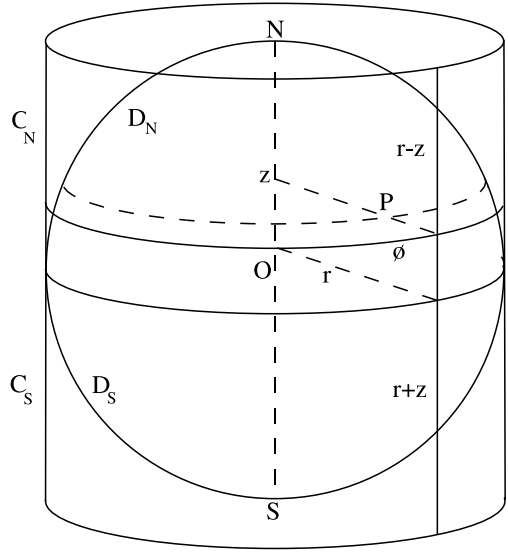}
\end{center}
\caption{The Archimedes map is area preserving.} 
\end{figure}

   Consider two spherical caps $D_{N}$ and $D_{S}$ which are 
determined by the parallel $z$ respectively around the north 
and the south poles. The images of $D_{N}$ and $D_{S}$ are 
respectively given by the cylindrical regions $C_{N}$ and 
$C_{S}$. The area of a very small region on the sphere is 
given by the $2$-form $s=\sin(\nu\theta)d\theta \wedge d\phi$.
The $1$-forms 

\begin{eqnarray}
a^{N}=\frac{1}{\nu} [1-\cos(\nu\theta)]d\phi \hspace*{5mm} , \hspace*{5mm}
a^{S}=-\frac{1}{\nu} [1+\cos(\nu\theta)]d\phi ,
\end{eqnarray}

\noindent respectively yield the area $2$-form $s^{N}=s^{S}=s$ 
over $D_{N}$ and $D_{S}$: $s^{N}=da^{N}$, $s^{S}=da^{S}$. The 
area of the corresponding region on the cylinder is given by 
$h=-dz \wedge d\phi$. Then $h^{N}=db^{N}$, $h^{S}=db^{S}$ where

\begin{eqnarray}
b^{N}=\left( \frac{1}{\nu} - z \right) d\phi \hspace*{5mm} , \hspace*{5mm}
b^{S}=- \left( \frac{1}{\nu} + z \right) d\phi . 
\end{eqnarray}

\noindent Here $b^{N}$ is the area of the vertical strip between 
the parallels $z=1/\nu$ and $z=(1/\nu)\cos(\nu\theta)$ of height
$k^{N}=(1/\nu)-z$ and base length $d\phi$. Meanwhile $-b^{S}$ is 
the area of the vertical strip between the parallels 
$z=(1/\nu)\cos(\nu\theta)$ and $z=-1/\nu$ of height $k^{S}=(1/\nu)+z$ 
and base $d\phi$. The area of $D_{N}$ and $D_{S}$

\begin{eqnarray} 
& & S^{N}= \int_{D_{N}} s^{N} = \oint_{\partial D_{N}:P[\theta]} a^{N}  
\hspace*{5mm} , \hspace*{5mm} 
S^{S}= \int_{D_{S}} s^{S} = - \oint_{\partial D_{S}:P[\theta]} a^{S}  
\nonumber \\ 
& & \hspace*{7mm} =2 \pi \frac{1}{\nu^{2}} \left[ 1-\cos(\nu\theta) \right] 
\hspace*{23mm}
=2 \pi \frac{1}{\nu^{2}} \left[ 1+\cos(\nu\theta) \right] \nonumber \\
& & \hspace*{7mm} = 2 \pi \frac{1}{\nu} \left( \frac{1}{\nu}-z \right) 
\hspace*{32mm}
=2 \pi \frac{1}{\nu} \left( \frac{1}{\nu}+z \right) ,
\end{eqnarray}

\noindent are respectively equal to the area of the regions 
$C_{N}$ and $C_{S}$. These are respectively given by the area 
of the rectangles with base $2 \pi (1/\nu)$ and height 
$k^{N}$ and $k^{S}$.

   The Wu-Yang solution is based on pasting the caps  $D_{N}$ 
and $D_{S}$ along the common boundary $P: z(\theta)$. We need a 
dimensionful factor: $\nu^{2}/2g$ 
for the sake of dimensional consistency: $F=(\nu^{2}/2g) s$, 
$A^{N}=(\nu^{2}/2g) a^{N}$, $A^{S}=(\nu^{2}/2g) a^{S}$ and 
$Q^{N}=(\nu^{2}/2g) S^{N}$, $Q^{S}=(\nu^{2}/2g) S^{S}$.
The gauge transformation formula (\ref{WuYanggaugetransformation}) 
yields $U=\exp [i(2\phi/\nu)\nu^{2}/2]$ where the factor $2\phi/\nu$ 
is the area of a vertical strip of height $2/\nu$ and base $\phi$ on 
the cylinder. This leads to $U=\exp (igQ)$ where 
$Q= (4\pi/\nu^{2})\nu^{2}/2g$, in terms of the total area of the 
sphere or the cylinder. The $1$-forms $*F^{N}$ and $*F^{S}$ are given 
as: $*F^{N}=\nu A^{N}$, $*F^{S}=\nu A^{S}$. These yield 
$\Phi^{N}=\pi(\nu^{2}/g) k^{N}$, $\Phi^{S}=\pi(\nu^{2}/g) k^{S}$.
Thus we can interpret the integrated field equations (\ref{fluxcharge}) 
simply as the area formula: $height=(1/base)area$ of a rectangle of height 
$k$ and base $2\pi/\nu$ up to an overall factor of $\nu^{2}/2g$  

  The quantization of the topological mass: $\nu=ng^{2}$ can 
be interpreted as division of the rectangle with base 
$2\pi/g^{2}$ and height $2/g^{2}$ into $n^{2}$ 
similar sub-rectangles of equal area. 

\subsection{The Geometric Phase}

  The analogy between the Berry phase \cite{Berry}, which 
corresponds to holonomy in a line bundle \cite{Simon}, and 
the magnetic monopole is well known, \cite{Zumino, Kritsis}, 
\cite{Hannay, AnanStod, Li, BFH}, \cite{A}, \cite{Mostafazadeh}.
The classical example of this is the geometric phase suffered 
by a vector upon parallel transport along the lattitude $\theta$  
on the $2$-sphere \cite{AF}, \cite{Nakahara}. We can express this 
geometric phase in terms of holonomy of the topologically massive 
gauge potential or the dual-field.

  Consider a vector $X$ which is tangent to $S^{2}$ at longitude 
$\phi=0$ and latitude $\theta$ as shown in Figure 5. If we parallel 
transport this vector along the latitude, after a complete revolution, 
it suffers a phase $\gamma$ which is determined by the solid angle 
$\Omega$ subtented by the parallel $P$ at latitude $\theta$ in $S^{2}$ 
\cite{Berry}, \cite{Hannay, AnanStod, Li, BFH}, \cite{Mostafazadeh}. 
This is given as

\begin{eqnarray} \label{Solidangle}
& & \gamma_{N}=-\Omega_{N} 
\hspace*{25mm} , \hspace*{15mm}
\gamma_{S}=\Omega_{S} \\
& & \hspace*{6mm} =-2\pi [1-\cos(\nu\theta)] 
\hspace*{25mm} 
=2\pi [1+\cos(\nu\theta)], \nonumber 
\end{eqnarray}

\noindent using the arclength parameterization. Note that 
$\gamma_{N}-\gamma_{S}=-4\pi$ although $\gamma_{N} \neq \gamma_{S}$. 
It is straightforward to verify this by solving the equation for parallel 
transport $\nabla_{(\phi)}X=0$ with the metric (\ref{defectedS2metric}) on 
$S^{2}$ \cite{Nakahara}. This equation reduces to

\begin{eqnarray}
\frac{dZ}{d\phi}+i\nu\cos(\nu\theta)Z=0,
\end{eqnarray}

\noindent \cite{AF}. We find 

\begin{eqnarray}\label{solution}
Z(2\pi\frac{1}{\nu})=Z(0)\exp\{-i\oint_{P}\nu[1-\cos(\nu\theta)]d\phi\} ,
\end{eqnarray}

\noindent for a complete revolution. Here $Z(0)$ and $Z(2\pi/\nu)$ 
respectively correspond to the initial and the final vectors $X_{i}$, 
$X_{f}$. The phase in (\ref{solution}) reduces to the solid angle in 
(\ref{Solidangle}) which is subtented by the parallel $P$ via the Stokes 
theorem. 

\newpage

\begin{figure}[!ht]
\begin{center}
\includegraphics[scale=0.4]{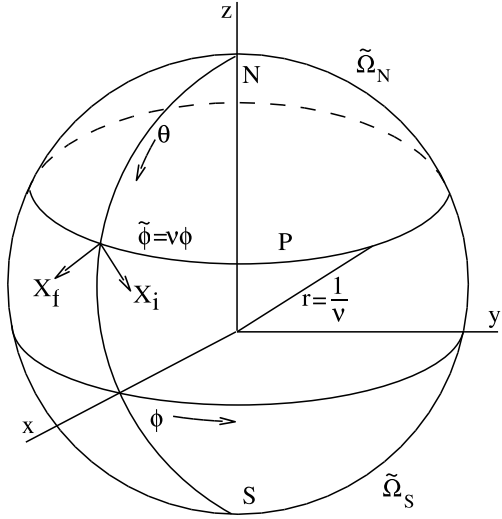}
\end{center}
\caption{The geometric phase $\tilde{\gamma}$ suffered 
by a vector $X$ upon parallel transport along the latitude 
$P: \theta$ is determined by the solid angle $\tilde{\Omega}$.} 
\end{figure}

  The phases (\ref{Solidangle}) can be written in terms of holonomy 
$\Gamma$ of the topologically massive gauge potential or the dual-field 
over the parallel $P$ as

\begin{eqnarray}
\gamma_{N}=-2\Gamma_{N} \hspace*{15mm} , \hspace*{15mm}
\gamma_{S}=-2\Gamma_{S},
\end{eqnarray}

\noindent where 

\begin{eqnarray} \label{Holonomy}
\Gamma_{N}=Q^{N}=\frac{1}{\nu}\Phi^{N} 
\hspace*{15mm} , \hspace*{15mm}
\Gamma_{S}=-Q^{S}=-\frac{1}{\nu}\Phi^{S} ,
\end{eqnarray}

\noindent (\ref{partialcharges}) and the factor $1/g$ is ignored. Here 
$\Gamma_{N}-\Gamma_{S}=2\pi$. The phase factors corresponding to holonomy 
of the sections $\tilde{A}^{N}$, $\tilde{A}^{S}$ of the potential 
(\ref{generalizedconnectionk}) are equal

\begin{eqnarray}
\exp(i\Gamma_{N})=\exp(i\Gamma_{S}),
\end{eqnarray}

\noindent since $p=1+k$ is an integer.

\section{The Quantization of Topological Mass}
 
  If the strength of the potentials $A^{N}$, $A^{S}$ 
(\ref{AbelianpotentialS2}) were $\pm(1/2)mg$ then the gauge function 
would be $U=\exp{(img^{2}\phi)}=\exp{(ip\nu\phi)}$, see section $3.5$. 
This would yield $U=\exp{(i2\pi mg^{2}/\nu)}$ in terms of the total 
magnetic and electric charges $\Phi=\nu Q=2\pi mg$.

  The expression $U=\exp{(i2\pi mg^{2}/\nu)}=\exp{(i2\pi p)}$ implies 
$\nu = ng^{2}$ with $p=\frac{m}{n}=1+k$ an integer. Hence $m$ is quantized 
in units of $n$. Here $p$ is the winding number of the gauge transformation 
which corresponds to the negative of the first Chern number. Thus the 
arclength parameterization leads to $r=1/ng^{2}$. Then $U=\exp{(img^{2}\phi)}$ 
is single-valued on this circle. Because as 
$\nu\phi \rightarrow \nu\phi + 2\pi$

\begin{eqnarray}
U=\exp{(img^{2}\phi)} \longrightarrow \exp{(i2\pi p)}\exp{(img^{2}\phi)}
\hspace*{3mm} , \hspace*{5mm} 
m=pn ,
\end{eqnarray}

\noindent since $0 \leq \nu \phi \leq 2\pi$. More generally,
$U \rightarrow \exp {(i2\pi pl)} U$ as $\nu\phi\rightarrow\nu\phi+2\pi l$ 
where $l$ is an integer. Then $l^{\prime}=pl$ is an integer too. Note that 
this discussion implicitly amounts to a redefinition of the arclength:
$\phi \rightarrow p\phi$ without changing the length scale defined by $\nu$.

   Further we require $U=\exp{(img^{2}\phi)}$ to be a single-valued
function of $\phi$ independent of any a priori choice of a scale of 
length, neither by redefining the arclength $\phi$ nor by imposing 
the condition $0 \leq \phi \leq L=2\pi/\nu$ on it. The identification 
$r=1/ng^{2}$ provides a fundamental unit of length $1/g^{2}$ or 
$\lambda=2\pi/g^{2}$ over which any gauge function is single-valued. 
Then as $g^{2} \phi \rightarrow g^{2} \phi + 2\pi$ 

\begin{eqnarray}
U=\exp{(img^{2}\phi)} \longrightarrow \exp{(i2\pi m)}\exp{(img^{2}\phi)} ,
\end{eqnarray}

\noindent since $0 \leq g^{2} \phi \leq 2\pi$. This yields an integer 
value for $m$. More generally, $U\rightarrow\exp{(i2\pi mk)}U$ as 
$g^{2}\phi \rightarrow  g^{2}\phi + 2\pi k$ where $k$ is an integer. 
Then $l^{\prime}=mk$ is an integer too. This leads to the quantization 
of the topological mass. Because if $m=pn$ is an integer for any integer 
$p$ then $n$ is an integer too. The  converse is also true.

  Thus the quantization of topological mass further requires that 
the circle of perimeter $L=2\pi/\nu$ has been wound $n$ times by the 
fundamental circle of perimeter $\lambda$ with locally invariant 
arclength. This leads to $l=nk$ which yields $l^{\prime}=pl=mk$. 
More precisely, as $\phi$ changes from $\phi=0$ to $\phi=lL=l(\lambda/n)$ 
this circle is traversed $l$ times. Meanwhile, the fundamental circle is 
traversed $k=l/n$ times. If $k$ is an integer, then $l=nk$ is an integer 
if and only if $n$ is an integer too. Thus the fundamental circle covers 
this an integer $(l=)n$ times and its radius is $r=\lambda/(2\pi n)=1/ng^{2}$. 
The quantization of topological mass amounts to a discrete change of scale 
which arises from a winding by the circle of fundamental size with invariant 
arclength. Intuitively number $n$ is a measure of the \textit{topological 
thickness} of the winding as the name \textit{topological mass} suggests. 

  We can see this if we rewrite $\nu=ng^{2}$ as

\begin{eqnarray} \label{DimBetterQuantTopMass}
\frac{\lambda}{{2\pi\frac{1}{\nu}}} = n .
\end{eqnarray}

\noindent Then the gauge function $U=\exp{(i\nu\phi)}$ is given as 

\begin{eqnarray} \label{GaugefunctionFourier}
U=\exp {(i\frac{2\pi n}{\lambda}\phi)} .
\end{eqnarray}

\noindent If $U$ is a single-valued function of $\phi$ with the 
fundamental length scale $\lambda$ for any $n$ then $n$ has to 
be an integer. The fundamental length scale $\lambda=nL$ is the 
least common multiple of intervals over which the gauge function 
is single-valued and periodic for any integer $n$ in addition to 
the fact that it has a smaller period $L=\lambda/n$ \cite{Pen}. 

  In other words the quantization of topological mass reduces to 
quantization of the inverse natural length scale in units of the 
inverse fundamental scale. We can associate the integer $n$ with a
winding number of mapping of circles using the arclength parameterization.

  In the fundamental units: $g=1$ the arclength $\phi$ reduces to angle: 
$0 \leq \phi \leq 2\pi$ and the topological mass becomes $\nu=n$. One could 
identify the number $n$ with the topological winding number $p$ ignoring 
the physical dimensions and discuss the quantization of topological mass. 
If $U=\exp{(i\nu\phi)}$ is single-valued on the unit circle then $n$ is 
an integer. 

\subsection{The Winding Numbers} 

  The Hopf map (\ref{Hopfmap}) defines a $S^{1}$ bundle over $S^{2}$ with 
a natural scale of length determined by the topological mass. The gauge 
function, that is the transition function of the bundle is given by a map 
$h: S^{1}_{(2)} \longrightarrow {S^{\prime}}^{1}_{(2)}$ from the equator 
$S^{1}_{(2)}= U_{N} \bigcap U_{S}$ of $S^{2}$ to the manifold of the gauge 
group ${S^{\prime}}^{1}_{(2)}$ \cite{AzcarragaIzquierdo}. The group $U(1)$ 
is given by normalizing the radius of ${S^{\prime}}^{1}_{(2)}$. Topological 
structure of the bundle is determined by homotopy class of the transition 
function which is characterized by the winding number of $S^{1}_{(2)}$ about 
${S^{\prime}}^{1}_{(2)}$ \cite{AzcarragaIzquierdo, Steenrod}. This winding 
number, which is associated with the magnetic charge of the monopole, 
corresponds to the negative of the First Chern number of the bundle. 
Meanwhile the size of ${S^{\prime}}^{1}_{(2)}$ (also $S^{1}_{(2)}$) is 
determined by a winding $f: S^{1}_{(1)} \longrightarrow {S^{\prime}}^{1}_{(2)}$
of the circle of fundamental size with locally invariant arclength. 

  The winding number of a map 
$u: \tilde{S}^{1}_{(1)} \longrightarrow  \tilde{S}^{1}_{(2)}$ 
is defined as the degree of this map

\begin{eqnarray} \label{degree}
\oint_{\tilde{S}^{1}_{(1)}} u^{*}(\omega_{(2)}) 
= deg(u) \oint_{\tilde{S}^{1}_{(2)}} \omega_{(2)} ,
\end{eqnarray}

\noindent \cite{GreuHalpVan, Felsager, MadTorn}. Here $\omega_{(2)}$ is 
the arclength $1$-form on $\tilde{S}^{1}_{(2)}$ and $u^{*}(\omega_{(2)})$ 
is its pull-back onto $\tilde{S}^{1}_{(1)}$. The map $u$ is uniquely 
defined for each point in $\tilde{S}^{1}_{(1)}$ and it is single-valued 
in $\tilde{S}^{1}_{(2)}$ if and only if the degree $deg(u)$, which 
measures the effective number $\tilde{S}^{1}_{(2)}$ is covered by 
$\tilde{S}^{1}_{(1)}$, is an integer. Note that this definition is 
independent of choice of a length scale. In the non-massive case, which 
has no natural scale, this map is defined only in terms of angles. In 
the present case we use arclength parameterization for defining this 
in accordance with the natural scale of length.

  The gauge function $U$ is given by
$h: S^{1}_{(2)} \longrightarrow {S^{\prime}}^{1}_{(2)}$ 

\begin{eqnarray}
\exp (i\nu\phi^{\prime}_{(2)})
=\exp (ip\nu\phi_{(2)}),
\end{eqnarray}

\noindent where the circles $S^{1}_{(2)}$ and ${S^{\prime}}^{1}_{(2)}$
are of the same radius $r^{\prime}_{(2)}=r_{(2)}=1/\nu$. Then 
$\phi^{\prime}_{(2)}=p\phi_{(2)}$ while 
$\tilde{\phi}^{\prime}_{(2)}=p\tilde{\phi}_{(2)}$. 
We find

\begin{eqnarray} \label{degreeh}
deg(h)=p, 
\end{eqnarray} 

\noindent since $r_{(2)}/r^{\prime}_{(2)}=1$ (\ref{degree}). Hence, 
as $S^{1}_{(2)}$ is traversed $l$ times, ${S^{\prime}}^{1}_{(2)}$ is 
traversed $l^{\prime}=pl$ times. Note that $deg(h)=p$ is independent 
of choice of any length scale $r^{\prime}_{(2)}=r_{(2)}$. 

   Consider the function $g: S^{1}_{(1)} \longrightarrow S^{1}_{(2)}$

\begin{eqnarray}
\hspace*{10mm} \exp (i\nu\phi_{(2)})=\exp (in\nu_{(1)}\phi_{(1)}), 
\end{eqnarray}

\noindent which maps the circle $S^{1}_{(1)}$ onto $S^{1}_{(2)}$ 
with locally invariant arclength $\phi_{(2)}=\phi_{(1)}$ where 
$\nu_{(1)}=1/g^{2}$. Then $r_{(1)}=nr_{(2)}$ and $\lambda=nL$ while 
$\tilde{\phi}_{(2)}=n\tilde{\phi}_{(1)}$. The degree formula 
(\ref{degree}) yields the winding number as the ratio $r_{(1)}/r_{(2)}$

\begin{eqnarray} \label{degreeg}
deg(g)=n. 
\end{eqnarray} 

\noindent Hence, as $S^{1}_{(1)}$ is traversed $k$ times, $S^{1}_{(2)}$ is 
traversed $l=nk$ times. The radius $r_{(2)}$ of circle $S^{1}_{(2)}$ is 
determined by the winding number $n=r_{(1)}/r_{(2)}$.

   The composite mapping $f=h \cdot g$ is given as

\begin{eqnarray}
\exp (i\nu\phi^{\prime}_{(2)})
=\exp (im\nu_{(1)}\phi_{(1)}).
\end{eqnarray}

\noindent In this case $\phi^{\prime}_{(2)}=p\phi_{(1)}$ 
while $\tilde{\phi}^{\prime}_{(2)}=m\tilde{\phi}_{(1)}$ 
since $r_{(1)}=nr^{\prime}_{(2)}$, $m=np$. The degree formula 
(\ref{degree}) yields the winding number as the product 

\begin{eqnarray} \label{degreef}
deg(f)=m.
\end{eqnarray} 

\noindent Then as $S^{1}_{(1)}$ is traversed $k$ times, 
${S^{\prime}}^{1}_{(2)}$ is traversed $l^{\prime}=mk$ times. 
This is equivalent to the case in $g$. That is, $m=np$ is an 
integer for any integer $p$ if and only if $n$ is an integer 
too. We find $m=n$ for $p=1$ while $m=p$ for $r_{(1)}/r_{(2)}=n=1$.

\subsection{Dirac Quantization Condition}

  We can now elucidate the acclaimed relation between the quantization of 
topological mass and Dirac quantization condition \cite{Polychronakos, HenTei}.
Consider Dirac potentials $A^{N/S}$ of \textit{magnetic strength} $g^{\prime}$ 
on the respective charts $U_{N/S}$ of $S^{2}$ with radius $r=1/\nu$

\begin{eqnarray}
A^{N/S}=\pm\frac{1}{2}g^{\prime} \nu 
\left[ 1\mp\cos(\nu\theta) \right] d\phi
=\pm\frac{1}{2} n^{\prime} \frac{\nu}{g}
\left[ 1\mp\cos(\nu\theta) \right] d\phi, 
\end{eqnarray}

\noindent where $gg^{\prime}=n^{\prime}$. The relations
$gg^{\prime}=n^{\prime}$ and $\nu=ng^{2}$ yield $\nu g^{\prime}=sg$ 
\cite{HenTei}, $s=nn^{\prime}$. The magnetic and electric charges 
associated with the potentials $A^{N/S}$ are

\begin{eqnarray}
Q^{N/S}=2\pi g^{\prime}, \hspace*{10mm} 
& & \Phi^{N/S}=2\pi g^{\prime} \nu=2\pi s g. 
\end{eqnarray}

  The potentials $A^{N/S}$ differ by a gauge transformation  
$U=\exp(igg^{\prime}\nu\phi)$. The Wu-Yang construction yields the 
gauge function as $U=\exp(i2\pi gg^{\prime})$ in terms of the magnetic 
or electric charges. Thus $gg^{\prime}=n^{\prime}$ or equivalently 
$p^{\prime}=n^{\prime}$ $(p=1)$ is an integer. If $m^{\prime}=s$
is to be an integer as described above then $n$ is an integer too.

\section{Conclusion}

   We have considered euclidean Wu-Yang type solutions of the MCS and 
the $SU(2)$ YMCS theories on the dS space $S^{3}$. We have discussed 
the introduction of a physical scale of length determined by the inverse 
topological mass $\nu \sim ng^{2}$ in these solutions from a geometrical 
point of view. We have embedded the abelian solution into the YMCS theory 
by means of a $SU(2)$ gauge transformation which has winding number $w=1$. 
The total action for this solution only consists of the YM piece. The action 
for the abelian solution vanishes.

  In the abelian case the field locally determines the potential up 
to a closed $1$-form via the self-duality equation. We have introduced 
a transformation of the gauge potential using the dual-field strength 
which can be identified with a gauge transformation.

  Then we have discussed the Hopf map $S^{3} \longrightarrow S^{2}$ 
including the topological mass. This has led to a reduction of the field 
equation onto $S^{2}$ using local sections of $S^{3}$. The Hopf map has 
given rise to the local potentials on $S^{2}$ which differ by a gauge 
transformation. These solutions have a composite structure consisting 
of both magnetic and electric charges because of the local character of 
the field equation. The quantization of the total electric charge $\Phi$ 
requires the quantization of the topological mass.

   This has naturally led to the Wu-Yang solution which is 
based on seaming the local potentials across the common boundary 
by means of a gauge transformation. The gauge function can be 
expressed in terms of either the magnetic or the electric charges. 
We have also presented abelian solutions with different first Chern 
numbers. 
 
  We have also presented three geometric structures which arise in 
connection with our solutions. We have presented a stereographic 
view of the fibers in the Hopf map. The Archimedes map has provided 
a simple picture for the Wu-Yang solution. This has led to a simple 
interpretation of the composite structure in terms of the area of a 
rectangle. We have also expressed the geometric phase suffered by a 
vector upon parallel transport on $S^{2}$ in terms of the holonomy 
of the topologically massive gauge potential or the dual-field. 

  Then we have discussed quantization of the topological mass $\nu=ng^{2}$ 
in the abelian case. In geometric terms, quantization of the topological 
mass reduces to quantization of inverse of the natural scale of length 
$L=2\pi/ng^{2}$ in units of the inverse of the fundamental length scale 
$\lambda=2\pi/g^{2}$. The fundamental length scale $\lambda$ is the least 
common multiple of intervals over which the gauge function is single-valued 
and periodic for any integer $n$ in addition to the fact that it has a smaller 
period $L=\lambda/n$. The number $n$ can be identified with the winding
number of a function that maps the circle of fundamental size about this 
circle with locally invariant arclength. Meanwhile the First Chern number
is given by the winding number of maps of circles of the same size $L$. 
We have also discussed Dirac quantization condition.

  Finally, we would like to point out a naive analogy between the natural 
length scale introduced by the topological mass and the Hall resistivity 
$R=1/ng^{2}$. The monopole/instanton type solutions presented here might 
provide a geometrical explanation for the levels if the \textit{particles} 
in the Hall effect gave rise to a gauge theoretic system with a fundamental 
length scale $1/g^{2}$. An immediate prediction would be the effect of 
quenching if the sample size is of the order of radius of these particles'
orbits, while the smaller orbits survived.

  We remark that the most distinctive feature of the topologically massive
gauge theories is the existence of a natural scale of length introduced by 
the topological mass. We have demonstrated that incorporation of this crucial 
fact which was ignored in the literature, settles down the physical discussions
into a consistent geometric frame. To the knowledge of the author, the 
solutions and geometric constructions presented here which clarify this 
issue are lacking in the literature.

\end{document}